\documentclass[preprint,12pt]{elsarticle}
\usepackage[colorlinks]{hyperref}
\usepackage{amssymb}
\usepackage{amsthm}
\usepackage{graphics}
\usepackage{epsfig}

\biboptions{sort&compress}

\begin{document}

\begin{frontmatter}

\title{Spontaneously induced general relativity with holographic interior
and general exterior}

\author[add1]{Xiao-Qing Shen}

\author[add1]{Shao-Feng Wu\corref{cor1}}
\ead{sfwu@shu.edu.cn}
\author[add1]{Guo-Hong Yang}

\cortext[cor1]{Corresponding author}
\address[add1]{Department of Physics, Shanghai University, Shanghai 200444, P.R. China}

\journal{Physics Letters B}
\begin{abstract}
We study the spontaneously induced general relativity (GR) from the
scalar-tensor gravity. We demonstrate by numerical methods that a
novel inner core can be connected to the Schwarzschild exterior with
cosmological constants and any sectional curvature. Deriving an
analytic core metric for a general exterior, we show that all the
nontrivial features of the core, including the locally holographic
entropy packing, are universal for the general exterior in static
spacetimes. We also investigate whether the f(R) gravity can
accommodate the nontrivial core.
\end{abstract}

\end{frontmatter}

\section{Introduction}

Recently, by studying the scalar-tensor gravity, Davidson and Gurwich \cite%
{Davidson1} found an interesting metric, for which the exterior
Schwarzschild solution of GR is spontaneously induced by a phase transition
which occurs precisely at the would-have-been event horizon (see earlier
discussion on the horizon phase transition \cite{Davidson2}), and the
interior core is characterized by some nontrivial features. For instance,
the core has a vanishing spatial volume. This provides a simple but direct
explanation of the well-known problem: why the BH entropy is not
proportional to the volume of the system as usual. Note that the volume of
normal BH can not be well-defined \cite{Brandon}, partially because the r$%
\leftrightarrow $t signature flip, but in the current situation, it makes
better since no such flip. Moreover, compared with the Schwarzschild BH, the
Komar mass \cite{Komar} within the core is well defined (non-singular and
positive). Then the Smarr formula \cite{Smarr} can be extended to any inner
sphere on which the area-entropy relation is extracted. It was argued that
this provides a local realization of maximal entropy packing and a new way
of Nature's ultimate information storage. In Ref. \cite{Davidson3}, Davidson
further designed a discrete holographic shell model to capture the
holographic entropy packing inside BHs. It is interesting to see that even
the logarithmic correction can be recovered.

The Davidson-Gurwich (DG) horizon phase transition has been also found in
the case with the Reissner-Nordstrom (RN) exterior \cite{Davidson4}. A
natural problem is whether the phase transition can exist for a more general
exterior. Respecting the celebrated holographic correspondence between the
supergravity in Anti-de Sitter (AdS) space and conformal field theory (CFT)
on the boundary \cite{Maldacena1998}, one would most like to ask whether the
locally holographic entropy packing can be present inside Schwarzschild-AdS
BHs. Moreover, one may also concern about the realization of the DG horizon
phase transition in the asymptotically dS exterior, since the accelerated
expansion of present universe could be derived by the positive cosmological
constant and the earlier universe is well represented by dS-like
exponentially inflationary phase.

In this paper, by choosing a special scalar potential, we will demonstrate
the DG horizon phase transition with the Schwarzschild-dS/AdS exterior by
numerical methods. With the mind that the AdS/CFT correspondence usually
focuses on the CFT in the flat spacetime, we will consider the spacetime
with any sectional curvature. Furthermore, we will derive a general form of
the core metric by a new (semi-)analytic method and check the nontrivial
features.

On the other hand, in Refs. \cite{Davidson1,Davidson4}, it was suggested
that the DG phase transition can also be found in a simple theory of $f(R)$
gravity with the square curvature correction. However, we will show that
there is a subtle problem in the suggestion, that is, a precondition to the
equivalence between the scalar-tensor gravity and $f(R)$ gravity is not
satisfied directly for the core metric.

\section{A special scalar-tensor gravity}

Consider the scalar-tensor gravity without the kinetic term%
\begin{equation}
S=\frac{1}{16\pi }\int \left[ \phi R-V(\phi )\right] \sqrt{-g}~d^{4}x+S_{M}.
\label{st-action}
\end{equation}%
The gravitational field equations and the equation of motion for the scalar
field can be derived by varying this action with respect to $g_{\mu \nu }$
and $\phi $:%
\begin{equation}
\phi G_{ab}-\nabla _{a}\nabla _{b}\phi +g_{ab}\nabla ^{c}\nabla _{c}\phi +%
\frac{1}{2}g_{ab}V(\phi )-8\pi T_{ab}=0,  \label{stfeqs}
\end{equation}%
\begin{equation}
R-\frac{dV}{d\phi }=0,  \label{stseq}
\end{equation}%
where $T_{ab}$ denotes the energy-momentum tensor of matter.

Refs. \cite{Davidson1,Davidson4} consider the 4-dimensional static spacetime
with the spherical symmetry. In this paper, we will extend it to other
topological cases, which is interesting especially in the asymptotically AdS
space. We write the line element as%
\begin{equation}
ds^{2}=-e^{\nu (r)}dt^{2}+e^{\lambda (r)}dr^{2}+r^{2}d\Omega ^{2},
\label{stgle}
\end{equation}%
where $\Omega $ expresses the surface with the constant sectional curvature $%
k=0,\pm 1$.

To be consistent with the GR outside the core, the potential $V(\phi )$ has
been chosen as \cite{Davidson1,Davidson4}%
\begin{equation}
V(\phi )=\frac{3}{2a}\left( \phi -\frac{1}{G}\right) ^{2},  \label{V1}
\end{equation}%
where the value of $a$ can be made as small as necessary to be compatible
with Solar System tests. But for spontaneously inducing the GR with the
cosmological constant $\Lambda $, we will modify the potential to%
\begin{equation}
V(\phi )=\frac{3}{2a}\left( \phi -\frac{1}{G}\right) ^{2}+4\Lambda \left(
\phi -\frac{1}{G}\right) +\frac{2\Lambda }{G}.  \label{V}
\end{equation}%
Note that the second term in the RHS of Eq. (\ref{V}) is added so that Eq. (%
\ref{stseq}) can be equivalent to the contracted Einstein equation $%
R-4\Lambda =0$ when $\phi \rightarrow 1/G$ outside the core.

\section{Horizon phase transition: numerical method}

To obtain the master equations that having a nontrivial solution, one needs
to solve $\nu ^{\prime \prime }$ (the prime denotes the derivative with
respect to $r$) from Eq. (\ref{stseq})%
\begin{equation}
\nu ^{\prime \prime }=\frac{1}{2r^{2}}\left[ -4+4ke^{\lambda }-2e^{\lambda
}r^{2}\frac{dV}{d\phi }+r\left( \lambda ^{\prime }-\nu ^{\prime }\right)
(4+\nu ^{\prime })\right] ,  \label{vpp}
\end{equation}%
and use it to cancel $\nu ^{\prime \prime }$ in the field equation (\ref%
{stfeqs}). Then the independent components of field equations can be
reorganized as%
\begin{equation}
\phi ^{\prime \prime }-\frac{1}{2}\left( \nu ^{\prime }+\lambda ^{\prime
}\right) \phi ^{\prime }-\frac{1}{r}\left( \nu ^{\prime }+\lambda ^{\prime
}\right) \phi =0,  \label{eq1}
\end{equation}%
\begin{equation}
\phi ^{\prime \prime }+\frac{1}{2}\left( \nu ^{\prime }-\lambda ^{\prime
}\right) \left( \phi ^{\prime }-\frac{2}{r}\phi \right) -\frac{2}{r^{2}}%
\left( 1-ke^{\lambda }\right) \phi -\frac{2}{3}e^{\lambda }\left( \phi \frac{%
dV}{d\phi }-\frac{1}{2}V\right) =0,  \label{eq2}
\end{equation}%
\begin{equation}
\phi ^{\prime \prime }+\left[ \frac{1}{2}\left( \nu ^{\prime }-\lambda
^{\prime }\right) +\frac{2}{r}\right] \phi ^{\prime }-\frac{1}{3}e^{\lambda
}\left( \phi \frac{dV}{d\phi }-2V\right) =0.  \label{eq3}
\end{equation}%
These three equations with the potential (\ref{V}) are the master equations
that we will solve. Here we have set $T_{b}^{a}=0$ for our aim.

Let us determine the boundary conditions of master equations. Consider the
perturbation around a general exterior%
\[
\phi (r)=1+s\phi _{1}(r),\;\lambda (r)=-\log [g(r)+sL_{1}(r)],\;\nu (r)=\log
[h(r)+sN_{1}(r)],
\]%
where $g$ and $h$ denote the metric components of general GR exterior, and
the constant $s$ serves as a small expansion parameter. Substituting these
perturbative solutions into Eq. (\ref{eq3}), one can find a decoupled linear
differential equation at leading order%
\[
\phi _{1}^{\prime \prime }+\frac{1}{2}\left( \frac{4}{r}+\frac{g^{\prime }}{g%
}+\frac{h^{\prime }}{h}\right) \phi _{1}^{\prime }-\frac{1}{g}(\frac{1}{a}-%
\frac{4}{3}\Lambda )\phi _{1}=0.
\]%
Now we specify the exterior as%
\[
g=h=k-\frac{2M}{r}\pm \frac{r^{2}}{L^{2}},
\]%
where we have set $G=1$ and $\Lambda =\mp 3/L^{2}$ for convenience. Note
that the above/below branch refer to Schwarzschild-AdS/dS spacetimes. At
large $r$, we get rid of the diverging term to stay with the converging tail%
\begin{equation}
\phi _{1}\simeq r^{-\frac{3}{2}-\frac{5}{2}\sqrt{1\pm \frac{4L^{2}}{25a}}}.
\label{f1}
\end{equation}%
Obviously, the small parameter $a$ should be positive/negative for
negative/positive $\Lambda $. With the aid of $\phi _{1}$, one can obtain $%
L_{1}$ and $N_{1}$ from the remained master equations (\ref{eq1}) and (\ref%
{eq2}):%
\begin{equation}
L_{1}\simeq \pm \frac{5}{2L^{2}}\left( 1+\sqrt{1\pm \frac{4L^{2}}{25a}}%
\right) r^{\frac{1}{2}\left( 1-5\sqrt{1\pm \frac{4L^{2}}{25a}}\right) },
\label{L1}
\end{equation}%
\begin{equation}
N_{1}\simeq -\frac{5r}{2}\left( k-\frac{2M}{r}\right) \left( 1+\sqrt{1\pm
\frac{4L^{2}}{25a}}\right) r^{-\frac{5}{2}\left( 1+\sqrt{1\pm \frac{4L^{2}}{%
25a}}\right) },  \label{N1}
\end{equation}%
One should be careful that the term$\sim 2M/r$ in Eq. (\ref{N1}) is
necessary for the case with $k=0$.

\begin{figure}[tbp]
\includegraphics[width=13cm]{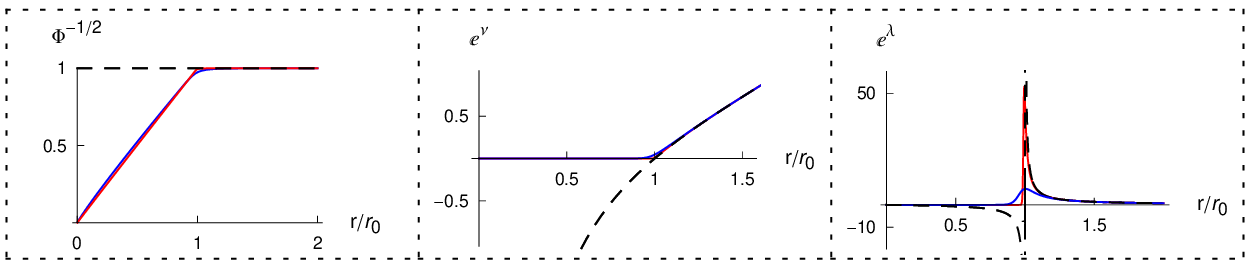}\newline
\caption{The configuration $e^{\protect\lambda }$, $e^{\protect\nu }$ and $%
\protect\phi $ with $\Lambda <0$ and $k=1$. The parameter in scalar
potential and the AdS radius are fixed as $a=0.1$ and $L=2$. The blue and
red lines refer to the expansion parameters $s=0.1$ and $0.01$. The dashed
lines depict the usual BH solutions with $s=0$.}
\label{fig1}
\end{figure}
\begin{figure}[tbp]
\includegraphics[width=13cm]{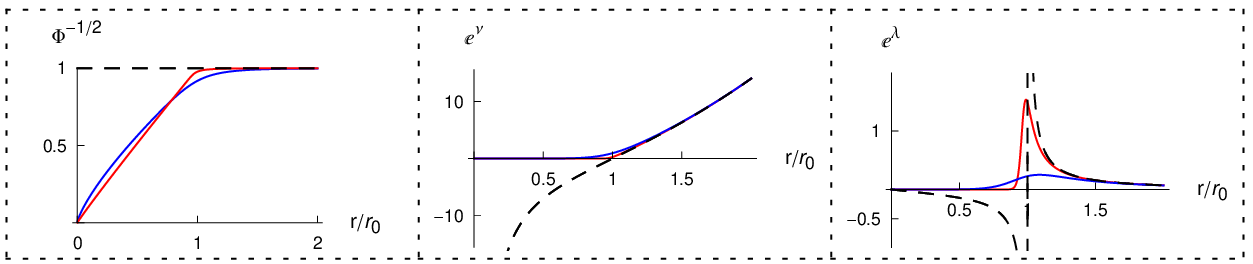}\newline
\caption{The configuration $e^{\protect\lambda }$, $e^{\protect\nu }$ and $%
\protect\phi $ with $\Lambda <0$ and $k=0$. The parameters are set as $%
a=0.05 $ and $L=1$. The blue and red lines refer to the expansion parameters
$s=0.1$ and $0.01$. The dashed lines depict the usual BH solutions with $s=0$%
.}
\label{fig2}
\end{figure}
\begin{figure}[tbp]
\includegraphics[width=13cm]{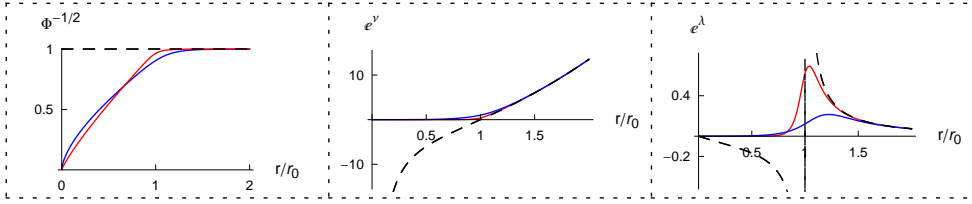}\newline
\caption{The configuration $e^{\protect\lambda }$, $e^{\protect\nu }$ and $%
\protect\phi $ with $\Lambda <0$ and $k=-1$. The parameters are set as $%
a=0.01$ and $L=0.5$. The blue and red lines refer to the expansion
parameters $s=0.1$ and $0.01$. The dashed lines depict the usual BH
solutions with $s=0$.}
\label{fig3}
\end{figure}

\begin{figure}[tbp]
\includegraphics[width=13cm]{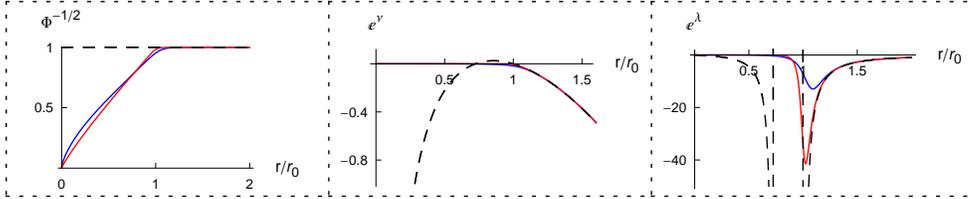}\newline
\caption{The configuration $e^{\protect\lambda }$, $e^{\protect\nu }$ and $%
\protect\phi $ with $\Lambda >0$. The parameter in scalar potential and the
dS radius are set as $a=-0.1$ and $L=1.5$. The blue and red lines refer to
the expansion parameters $s=10^{-3}$ and $10^{-4}$. The dashed lines depict
the usual BH solutions with $s=0$.}
\label{fig4}
\end{figure}

Hereto, we can perform a full numerical integration. For convention, the
horizon radius has been normalized to unity. Using Eqs. (\ref{f1}), (\ref{L1}%
) and (\ref{N1}) at some large enough distance as the boundary conditions,
we produce Figs. \ref{fig1}, \ref{fig2} and \ref{fig3} for the configuration
of $e^{\lambda }$, $e^{\nu }$ and $\phi $ with negative $\Lambda $ and
different $k$, as well as Fig. \ref{fig4} for the case of positive $\Lambda $%
. One can find that the exterior of topological-Schwarzschild-AdS BHs and
Schwarzschild-dS is recovered at the $s\rightarrow 0$ limit, and the inner
core differs conceptually from the interior of usual BHs with $s=0$ (see the
dashed lines). We like to stress three characteristic features of the
overall configuration: 1) $e^{\lambda }$ and $e^{\nu }$\ are drastically
suppressed inside the core; 2) there is no signature flip at the
would-have-been horizon; and 3) three functions $e^{\lambda }$, $e^{\nu }$
and $\phi $ at the $s\rightarrow 0$ limit are all changed abruptly very near
the would-have-been horizon, which manifests the horizon phase transition.

Moreover, one should be noted that the would-have-been horizon that connects
the Schwarzschild-dS exterior is located at the cosmological horizon,
instead of the event horizon in the cases of RN and Schwarzschild-AdS
exterior. From Fig. (\ref{fig4}), one can conclude that the total spacetime
is dynamical since $t$ denotes the spacelike coordinate and $r$ is timelike.

\section{Analytic inner core}

It this section, we will solve the master equations by a new analytic
method. One can find that it is consistent with the numerical results and it
is convenient to be used to check the nontrivial features of core in the
next section.

\subsection{General core metric with undetermined constants}

By numerical inspection, we have shown that $e^{\lambda }$ is negligible
inside the would-have-been horizon (see the first feature that we have
stressed in above section). This suggests us to simplify the master
equations by imposing $e^{\lambda }\rightarrow 0$ inside the core. With this
in mind, the master equations can be translated into%
\[
\phi ^{\prime \prime }=\phi ^{\prime }\left( -\frac{1}{r}+\frac{\phi
^{\prime }}{\phi }\right) ,\;\lambda ^{\prime }=\frac{2\left( \phi
^{2}+r\phi \phi ^{\prime }+r^{2}\phi ^{\prime 2}\right) }{r\phi \left( 2\phi
+r\phi ^{\prime }\right) },\;\nu ^{\prime }=-\frac{2\left( \phi +2r\phi
^{\prime }\right) }{r\left( 2\phi +r\phi ^{\prime }\right) }.
\]%
The analytic solution can be obtained as follows%
\begin{equation}
\phi =\gamma \left( \frac{r}{r_{0}}\right) ^{-2+\epsilon },\;e^{\lambda
}=\beta \left( \frac{r}{r_{0}}\right) ^{\frac{6}{\epsilon }-6+2\epsilon
},\;e^{\nu }=\alpha \left( \frac{r}{r_{0}}\right) ^{\frac{6}{\epsilon }-4},
\label{stcs}
\end{equation}%
where $\alpha $, $\beta $, $\gamma $ and $\epsilon $ are the constants of
integration. The scalar $r_{0}$ can be interpreted as the radius of the core
because the core metric (\ref{stcs}) can be self-consistent with the
condition $e^{\lambda }\rightarrow 0$ for all $r<r_{0}$, if one imposes $%
0<\epsilon <<1$. However, the scale $r_{0}$ is fictitious at this stage
since it can be absorbed into the coefficients $\alpha $, $\beta $, $\gamma $%
. The arbitrariness will be removed while we match the core metric with the
exterior solution.

The form of core metric (\ref{stcs}) with $k=1$ has been obtained for the
matter being vacuum \cite{Davidson1} or electromagnetic field \cite%
{Davidson4}. Here we have shown that the core metric can be derived with any
$k$ and without invoking the concrete scalar potential. Moreover, we note
that Eq. (\ref{stcs}) also can be derived using the equation (\ref{vpp}) and
the field equations with general matter content, provided that the matter
satisfies $e^{\lambda }T_{b}^{a}\rightarrow 0$ inside the core.

\subsection{Matching the core with a general exterior}

In Ref. \cite{Davidson4}, an analytic method was proposed to fix the
integration constants. This method solves field equations near the region of
phase transition then matches the result to the core and the RN exterior,
respectively. However, it seems difficult to extend the analytic method to
more general exterior. On the other hand, we note that there is a more
simple analytic method to solve differential equations, which directly
matches two approximate solutions near two boundaries. Recently, this
analytic method was introduced to understand the holographic superconductor
\cite{Gregory,Pan,Ge}. For the case of $d\leq 5$ dimensional spacetimes, the
analytic method can explain the qualitative features of superconductors and
gives fairly good agreement with numerical results. A natural idea is to use
this method to connect the core with the general exterior. Before doing
that, we would like to introduce a problem about the selection of the
matching point. In Ref. \cite{Gregory}, the matching point is selected as
the intermediate value $z_{m}=1/2$ between the range ($0<z<1$), but in fact,
it could be rather arbitrary without changing the qualitative features. For
the case of $d>5$, it was found\ \cite{Pan} that the matching point should
be chosen in an appropriate range ($z_{md}<z_{m}<1$), otherwise the behavior
of holographic superconductor can not be imitated qualitatively. In our
understanding, this means that the behavior of the exact solution is not
smooth enough to be simulated by the direct matching in the whole range.
Hence, one can suspect that for some differential equations which have very
sharp peaks, the appropriate range of matching points is very small, and
even no appropriate matching point can be found. In the following, one can
find that we are just encountering both cases. Fortunately, we still can
obtain correct results by a reasonable assumption.

At first, let us consider the matching of core metric $e^{\nu }$ and the
general GR solution $h$. For convention, we will change the variable $r$ to $%
z=r_{0}/r$, thus $z=1$ means the would-have-been horizon. The smooth
matching usually requires two functions at two boundaries and their
derivatives can be connected at one point:%
\begin{equation}
\left. \alpha \left( \frac{1}{z}\right) ^{\frac{6}{\epsilon }-4}\right\vert
_{z_{m}}=h(z_{m}),\;\left. \partial _{z}\left[ \alpha \left( \frac{1}{z}%
\right) ^{\frac{6}{\epsilon }-4}\right] \right\vert _{z_{m}}=\left. \partial
_{z}h(z)\right\vert _{z_{m}},  \label{h1}
\end{equation}%
which can be solved as%
\begin{equation}
h(z_{m})=\left. \frac{z\epsilon h^{\prime }}{-6+4\epsilon }\right\vert
_{z_{m}},  \label{ae1}
\end{equation}%
\begin{equation}
\alpha =\left. \frac{\left( \frac{1}{z}\right) ^{3-\frac{6}{\epsilon }%
}\epsilon h^{\prime }}{-6+4\epsilon }\right\vert _{z_{m}}.  \label{ae2}
\end{equation}%
From Eq. (\ref{ae1}) and Eq. (\ref{ae2}), one can find that $h(z)$ should
satisfy $\left\vert h(z_{m})\right\vert \ll 1$, $h^{\prime
}(z_{m})/h(z_{m})<0$, and $\alpha /h(z_{m})>0$, because $\epsilon $ should
be a very small positive number to suppress $e^{\lambda }$ and $e^{\nu }$,
and there is no signature flip at the would-have-been horizon (see the first
and second features which we have stressed in above section). This further
means that the matching point is restricted at a very small range of $%
z_{m}\lesssim 1$. Expanding Eq. (\ref{ae1}) near the horizon, we have%
\begin{equation}
z_{m}=\frac{6-5\epsilon }{6-4\epsilon }+\mathcal{O}(z_{m}-1)^{2}.
\label{zmh}
\end{equation}%
Substituting Eq. (\ref{zmh}) into Eq. (\ref{ae2}) and using $0<\epsilon <<1$%
, we obtain%
\begin{equation}
\alpha =-\frac{h^{\prime }(1)}{6e}\epsilon +\mathcal{O}(\epsilon )^{2},
\label{a}
\end{equation}%
where $e$ is the base of the natural logarithm.

Next, we will consider the matching of core metric $e^{\lambda }$ and the
general GR solution $1/g$. We require the core metric and exterior solution
can be connected at certain point, which leads to%
\begin{equation}
\left. \frac{1}{\beta }\left( \frac{1}{z}\right) ^{6-\frac{6}{\epsilon }%
-2\epsilon }\right\vert _{z_{m}}=g(z_{m}).  \label{bg1}
\end{equation}%
Naively, one might also require their derivatives connecting, which means%
\begin{equation}
\left. \partial _{z}\left[ \frac{1}{\beta }\left( \frac{1}{z}\right) ^{6-%
\frac{6}{\epsilon }-2\epsilon }\right] \right\vert _{z_{m}}=\left. \partial
_{z}g(z)\right\vert _{z_{m}}.  \label{bg2}
\end{equation}%
Combing Eq. (\ref{bg1}) and Eq. (\ref{bg2}), one can obtain%
\begin{equation}
g(z_{m})=\left. \frac{z\epsilon g^{\prime }}{6-6\epsilon +2\epsilon ^{2}}%
\right\vert _{z_{m}},  \label{g1}
\end{equation}%
\begin{equation}
\beta =\left. \frac{2\left( \frac{1}{z}\right) ^{7-\frac{6}{\epsilon }%
-2\epsilon }(3-3\epsilon +\epsilon ^{2})}{\epsilon g^{\prime }}\right\vert
_{z_{m}}.  \label{b1}
\end{equation}%
If carrying out the similar analysis below Eq. (\ref{ae2}), one can find
that $g^{\prime }(z_{m})/g(z_{m})$ should be positive and the matching point
should be located at $z_{m}\lesssim 1$. However, this result is conflicted
with all the numerical solutions that we have known. Analyzing the analytic
core metric (\ref{stcs}) and Schwarzschild exterior for instance, one can
find that Eq. (\ref{bg1}) can be easily satisfied but Eq. (\ref{bg2}) can
not hold outside the horizon. This phenomenon is more clear if one considers
the direct matching between $\phi =\gamma \left( \frac{r}{r_{0}}\right)
^{-2+\epsilon }$ and the GR exterior $\phi =1$, since the derivative of $%
\gamma \left( \frac{r}{r_{0}}\right) ^{-2+\epsilon }$ can not be vanishing
unless $\gamma =0$. For fixing two integration constants $\beta $ and $%
\epsilon $, we would like to give up the smooth matching but expect to
replace Eq. (\ref{bg2}) with another new condition. Here we propose that $%
e^{\lambda }$, $e^{\nu }$ and $\phi $ can be matched at the same position{%
\footnote{%
In Refs. \cite{Gregory,Pan,Ge}, the two fields are also matched at the same
position. But whether it is necessary has not been studied.}}. This is a
natural ansatz since it implies that we are considering the phase transition
which happens at the same position (see the third feature that we have
stressed in above section). Thus, expanding Eq. (\ref{bg1}) near the horizon
and using Eq. (\ref{zmh}), we have%
\begin{equation}
\beta =-\frac{6}{g^{\prime }(1)e}\frac{1}{\epsilon }+\mathcal{O}(\epsilon
)^{0}.  \label{b}
\end{equation}%
Similarly, in terms of Eq. (\ref{zmh}) and%
\begin{equation}
\left. \gamma \left( \frac{1}{z}\right) ^{-2+\epsilon }\right\vert
_{z_{m}}=1,  \label{g2}
\end{equation}%
we can fix the integration constant as $\gamma \simeq 1$.

Several remarks are in order. Firstly, we stress that Eq. (\ref{a}), Eq. (%
\ref{b}) and $\gamma \simeq 1$ can be viable to a general exterior. The
information of general exterior manifests in $h^{\prime }(1)$ and $g^{\prime
}(1)$. Thus, in the next section, we can study the features of the core
connecting a general exterior, while not restricted in a concrete BH.
Secondly, the explicit expressions of $\alpha $, $\beta $ and $\gamma $
demonstrate that $r_{0}$ is the horizon radius indeed. Thirdly, we can
recover the results obtained in Refs. \cite{Davidson1,Davidson4} for the
Schwarzschild or RN exterior, up to a factor of the number $e$.
Interestingly, the number $e$ will not affect any properties of the core, as
we will show in next section. Fourthly, in the above, we treat three
functions ($h$, $g$ and $\phi $) as independent functions. In fact, with the
mind that there are only four constants of integration $\alpha $, $\beta $, $%
\gamma $ and $\epsilon $, one can not impose three functions and their
derivatives connecting at the same time in general. To obtain the desired
nontrivial core, one can select four conditions as we have done, namely,
Eqs. (\ref{ae1}), (\ref{ae2}), (\ref{bg1}), and (\ref{g2}). Nevertheless, we
still have treated these functions as independent ones, because it suggests
a simple but robust analytic method, which could be applicable to the case
where the undetermined functions of some differential equations have enough
constants of integration and the exact solutions do not satisfy all the
conditions of the direct smooth matching. The key point is to find the
reasonable alternative conditions.

At last, we write the general form of the core metric for any exterior
solution%
\begin{equation}
ds^{2}=-\frac{\epsilon r_{0}h^{\prime }(r_{0})}{6e}\left( \frac{r}{r_{0}}%
\right) ^{\frac{6}{\epsilon }-4}dt^{2}+\frac{6}{\epsilon r_{0}g^{\prime
}(r_{0})e}\left( \frac{r}{r_{0}}\right) ^{\frac{6}{\epsilon }-6+2\epsilon
}dr^{2}+r^{2}d\Omega ^{2},  \label{stlespec}
\end{equation}%
where we have recovered $r_{0}$ explicitly for later convention.

\section{The nontrivial core features}

In this section, we will check whether the different exterior, namely, the
factors $h^{\prime }(r_{0})$ and $g^{\prime }(r_{0})$, and the number $e$
that characterizes our analytic method, will affect the nontrivial features
of the core given in \cite{Davidson1,Davidson4}. Before doing that, it
should be noted that the approach to the nontrivial features requires a
static spacetime. For instance, the Killing vector will be involved, which
is not well defined in a dynamic spacetime. So we will take into account the
core with a general static exterior, which excludes the case of
Schwarzschild-dS exterior.

\subsection{Vanishing spatial volume}

Associated with any surface $\Omega $ of radius $r$, the spatial volume is%
\begin{equation}
V(r)=\Omega \int_{0}^{r}\sqrt{g_{rr}}r^{2}dr\simeq \sqrt{\frac{6\epsilon }{%
r_{0}g^{\prime }(r_{0})e}}\left( \frac{r}{r_{0}}\right) ^{\frac{3}{\epsilon }%
}\frac{\Omega }{3}r_{0}^{3},  \label{stiv}
\end{equation}%
where we have used $\Omega $ to express the unit area of the surface $\Omega
$. In contrast to the corresponding finite surface area $A(r)=\Omega r^{2}$,
it is obvious that the spatial volume tends to vanish at the $\epsilon
\rightarrow +0$ limit.

\subsection{Constant temperature}

The surface gravity of any surface inside the core can be calculated as%
\begin{equation}
\kappa =\frac{1}{2}e^{-\lambda }\sqrt{e^{\lambda +\nu }}\nu ^{\prime }=-%
\frac{1}{6}\left( \frac{r}{r_{0}}\right) ^{-\epsilon }(2\epsilon -3)\sqrt{%
g^{\prime }(r_{0})h^{\prime }(r_{0})}\simeq \frac{1}{2}\sqrt{g^{\prime
}(r_{0})h^{\prime }(r_{0})}.  \label{stsg}
\end{equation}%
One can find that $\kappa $ tends to be a constant from $r=r_{0}$ to the
origin at $r=0$. Moreover, this constant is exactly identified as $2\pi $
the Hawking temperature of usual BHs. This suggests that the core is under
thermodynamic equilibrium. The nontrivial feature of the core temperature
can also be seen from the Rindler structure of the core:%
\begin{equation}
ds_{in}^{2}=-\kappa ^{2}\eta ^{2}dt^{2}+d\eta ^{2}+\left( \frac{3\kappa \eta
^{2}}{\epsilon r_{0}}\right) ^{\frac{3}{\epsilon }}r_{0}^{2}d\Omega ^{2},
\label{rindler}
\end{equation}%
where the proper length coordinate is%
\begin{equation}
\eta (r)=\int \sqrt{\beta \left( \frac{r}{r_{0}}\right) ^{\frac{6}{\epsilon }%
-6+2\epsilon }}dr=\frac{\left( \frac{r}{r_{0}}\right) ^{-2+\frac{3}{\epsilon
}+\epsilon }\sqrt{\frac{6r_{0}}{\epsilon eg^{\prime }(r_{0})}}}{\left( \frac{%
3}{\epsilon }-2+\epsilon \right) }.  \label{plcdef}
\end{equation}%
Thus, Hawking's imaginary time periodicity can be recovered from Eq. (\ref%
{rindler}). But unlike in the original GR BH, the Euclidean origin
corresponds now to the center of $r=0$ rather than to $r=r_{0}$. By
calculating the curvature scalar at leading order, for instance,
\[
R^{\mu \nu }R_{\mu \nu }\simeq \left( \frac{r}{r_{0}}\right) ^{-\frac{12}{%
\epsilon }+8+4\epsilon }2e^{2}g^{\prime }(r_{0})^{2}+\frac{2k^{2}}{r^{4}},
\]%
one should be noted that the origin is always the singularity of the general
metric{\footnote{%
The factor $g^{\prime }(r_{0})$ can not be vanishing otherwise the core
metric (\ref{stlespec}) is divergent.}}. This problem was expected \cite%
{Davidson1,Davidson4} to be solved by involving a more complicated
Lagrangian or quantum effects.

\subsection{Positive-definite mass}

Consider the Komar mass \cite{Komar}%
\begin{equation}
m_{K}=\frac{1}{4\pi }\int_{\Sigma }d\Sigma _{\mu }\xi _{\nu }R^{\mu \nu },
\label{stkomar}
\end{equation}%
where $\xi $ is the Killing vector and $\Sigma $ is a spatial volume with
the boundary $\Omega $. One can prove that for any surface $\Omega $, it
satisfies a geometric relation%
\begin{equation}
m_{K}(r)=\frac{1}{4\pi }\kappa (r)A(r),  \label{smarr}
\end{equation}%
which can be reduced to the Smarr formula for usual BHs \cite{Smarr}. The
Komar mass is meaningless in the interior of usual BHs, where $\kappa (r)$
and $m_{K}(r)$ can be negative and even negative infinite \cite%
{Davidson1,Davidson4}. Interestingly, however, $m_{K}(r)$ of the core metric
(\ref{stlespec}) is finite and positive-definite as desired for the
gravitational mass.

\subsection{Freezing light ray}

One can see that the radial light rays obey the null geodesics%
\begin{equation}
r(t)=r_{in}e^{\pm t\left/ \tilde{t}\right. },\ \tilde{t}=\frac{6r_{0}}{%
\epsilon \sqrt{g^{\prime }(r_{0})h^{\prime }(r_{0})}}.  \label{stinnugeos}
\end{equation}%
Thus, for an observer at asymptotic distance, a light ray sent from some $%
r_{in}<r_{0}$ is very difficult to move even a small distance. In other
words, it indicates that each layer of the core acts as an event horizon
which freezes the light ray if one looks from the outside. It hence seems
reasonable to associate the entropy on any inner surface to describe the
lack of the information behind it.

\subsection{Holographic entropy packing}

With the mind that the surface gravity is constant inside the core, Eq. (\ref%
{smarr}) suggests that every concentric inner surface of invariant surface
area $A(r)$ carries a geometric fraction $A(r)/A(r_{0})$ of the total Komar
mass enclosed by the would-have-been outer horizon, which also hints the
entropy packing inside the would-have-been horizon. Furthermore, since the
surface gravity can be exactly identified as $2\pi $ the Hawking temperature
of usual BHs, one can rewrite the Smarr formula (\ref{smarr}) as%
\begin{equation}
m_{K}(r)=2T(r_{0})\left( S(r_{0})\frac{r^{2}}{r_{0}^{2}}\right) .
\label{sttherformu}
\end{equation}%
Note that there does not exist an analogous formula for the interior of
ordinary BHs. Taking Eq. (\ref{sttherformu}) as a significant thermodynamic
equality, one could regard the expression in the parenthesis in Eq. (\ref%
{sttherformu}) as the entropy stored within an arbitrary inner surface%
\begin{equation}
S(r)=S(r_{0})\frac{r^{2}}{r_{0}^{2}}=\frac{A(r)}{4}.  \label{stenais}
\end{equation}%
Interestingly, this is just the universal holographic entropy bound, and
what is remarkable is that the bound is locally saturated.

Hereto, none of the features of the core which we have checked are
influenced by the factors of the general exterior. Moreover, all the
features also have not been affected by the number $e$ appeared in the core
metric (\ref{stlespec}), which supports the effectiveness of our analytic
method.

\section{About the $f(R)$\ gravity}

In general, the $f(R)$ theory can be equivalent to the Brans--Dicke theory
with the vanishing Brans--Dicke parameter, if the condition%
\begin{equation}
\phi =f_{R}=\frac{df(R)}{dR},\;V=Rf_{R}-f  \label{con}
\end{equation}%
is satisfied \cite{Felice}. Therefore, in Refs. \cite{Davidson1,Davidson4},
it was suggested that the DG horizon phase transition can also be found in a
simple $f(R)$ gravity with the square curvature correction%
\begin{equation}
f=R+\frac{a}{6}R^{2},  \label{fR}
\end{equation}%
since Eq. (\ref{fR}) and Eq. (\ref{V1}) satisfy Eq. (\ref{con}). Actually,
the equation of motion of scalar field (\ref{stseq}) can be expressed as%
\[
R=\frac{3}{a}(\phi -1),
\]%
which is equivalent to the condition%
\begin{equation}
\phi =f_{R}=1+\frac{a}{3}R.  \label{condition}
\end{equation}

However, one should be careful that, according to the metric ansatz (\ref%
{stgle}), the field equations (\ref{stfeqs}) (then the master equations)
include three independent equations which are enough to determine three
undetermined functions $\lambda $, $\nu $, and $\phi $. Thus, in general,
the obtained solutions will not satisfy Eq. (\ref{condition}), which hence
must be taken as an additional constraint. In the following, we will show
both analytically and numerically that the solution with the DG horizon
phase transition does not satisfy Eq. (\ref{condition}) indeed.

\begin{figure}[tbp]
\includegraphics[width=7cm]{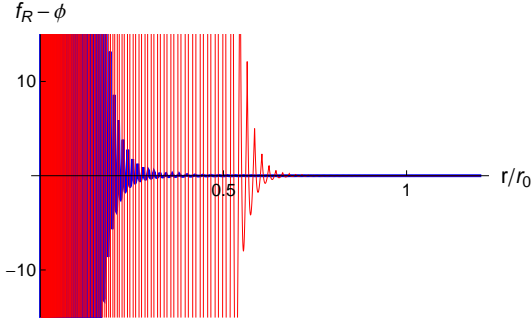}\newline
\caption{Check of the condition $f_{R}=\protect\phi $ for the core with
Schwarzschild exterior. The parameter $a=0.01$. The blue and red lines refer
to $s=0.1$ and $0.01$.}
\label{fig5}
\end{figure}

Using the analytic core metric (\ref{stlespec}) and $\phi \simeq \left(
\frac{r}{r_{0}}\right) ^{-2+\epsilon }$, one can calculate%
\begin{equation}
f_{R}-\phi =1+\frac{a}{3}R-\phi =1+\frac{2ka}{3r^{2}}-\left( \frac{r}{r_{0}}%
\right) ^{-2+\epsilon },  \label{ff}
\end{equation}%
which obviously can not be vanishing in general. Fig. \ref{fig5} gives the
numerical check of Eq. (\ref{condition}). One can find that $f_{R}-\phi $ is
not vanishing in general inside the core. In particular, it increases
rapidly when the parameter $s$ declines.

Thus, it seems difficult to argue that the DG horizon phase transition can
be found in the $f(R)$ gravity only because of the usual equivalence between
the scalar-tensor gravity and $f(R)$ gravity.

\section{Summary}

In this paper, we have studied the scalar-tensor gravity without the kinetic
term. Choosing a special scalar potential, we have shown by numerical
methods that there is the DG horizon phase transition, which connects the
topological-Schwarzschild-AdS or Schwarzschild-dS exterior with the inner
core that differs conceptually from the interior of usual BHs. We also find
an analytic expression of the core connecting a general exterior. We have
checked that the static analytic core has the nontrivial features, including
the vanishing spatial volume, constant temperature, positive-definitive
Komar mass, the freezing light ray, and the locally holographic entropy
packing. All these features are not changed by the factor of the general
exterior. Our results suggest that the spontaneously induced GR with
holographic interior could be viable for very general GR exterior in static
spacetimes.

Comparing the solutions with and without the cosmological constants, one can
find that the solution with Schwarzschild-dS exterior is very different with
the one with Schwarzschild or RN exterior, that is, the would-have-been
horizon that connects the Schwarzschild-dS exterior is located at the
cosmological horizon, instead of the event horizon. On the other hand, this
result is consistent with the solution with RN exterior, since the
would-have-been horizons are both the outer horizons. The positive
cosmological constant does not give the solution any qualitative difference.
We expect that the holographic core with asymptotically AdS exterior might
lead to some interesting results in the AdS/CFT correspondence.

At last, we have pointed out that the DG horizon phase transition might not
happen in the $f(R)$ gravity.

\emph{Acknowledgement:} This work was supported by NSFC (No. 10905037).

\end{document}